\documentclass[12pt]{revtex4}
\usepackage{amssymb}
\usepackage{graphicx}
\usepackage{epstopdf}

\def\be{\begin{equation}}
\def\eec{, \end{equation}}

\def\be{\begin{equation}}
\def\ee{ \end{equation}}

\newcommand{\lsim}{\raisebox{-.4ex}{$\stackrel{<}{\scriptstyle \sim}$}}
\newcommand{\gsim}{\raisebox{-.4ex}{$\stackrel{>}{\scriptstyle \sim}$}}

\begin{document}
\title{Observational evidences for spinning black holes: A proof of general relativity for
spacetime around rotating black holes}
\smallskip\smallskip 
\author{Banibrata Mukhopadhyay$^1$, Debbijoy Bhattacharya$^{2,3}$ and P. Sreekumar$^4$}
\affiliation{1. Department of Physics,  Indian Institute of
Science,  Bangalore 560012,  India}
\affiliation{2. Inter-University Centre for Astronomy and Astrophysics, Post Bag 4, 
Ganeshkhind, Pune 411007, India}
\affiliation{3. Manipal Centre for Natural Sciences,
Manipal University, Manipal 576104, India
}
\affiliation{4. Space Astronomy Group, ISRO Satellite Centre, Bangalore 560017, India}

\email{bm@physics.iisc.ernet.in, debbijoy.b@manipal.edu, pskumar@isac.gov.in}  

\vskip10cm 
\begin{abstract}
Since it was theorized by Kerr in 1963, determining the spin of
black holes from observed data was paid
very little attention until few years back. The main reasons behind
this were the unavailability of adequate data and the lack of
appropriate techniques. In this article, we explore determining/predicting
the spin of several black holes in X-ray binaries and in the center of
galaxies, using X-ray and gamma-ray satellite data. For X-ray binaries,
in order to explain observed quasi-periodic oscillations, 
our model predicts the spin parameter 
of underlying black holes. On the other hand, the nature of 
spin parameters of black holes in BL Lacs and 
Flat Spectrum Radio Quasars is predicted by studying the total luminosities
of systems based on Fermi $\gamma$-ray data. 
All sources considered here exhibit characteristics of spinning black holes,
which verifies natural existence of the Kerr metric.
\end{abstract}

\maketitle

  \\


\newpage
\section{Introduction}\label{intro}

The solution of Einstein equation for gravitational field of a spinning mass 
was obtained by Kerr \cite{kerr}, which was later generalized by Boyer \& Lindquist \cite{boyer}
to construct a maximal analytic extension of it. In the metric, the spin parameter `$a$' of
the gravitational object, i.e. black hole, is defined in such a way that $|a|\le M$, when $M$ is the mass 
of the black hole (speed of light ($c$) and Newton's gravitational constant ($G$) are 
chosen to be unity). Note that in the Boyer-Lindquist coordinate, the outer radius of the 
rotating, uncharged black hole is $r_+=M+\sqrt{M^2-a^2}$. Hence, for $|a|> M$, the collapsed 
object will form a naked singularity, rather than a black hole,
without an event horizon. In addition, for $a=0$, the solution reduces to the 
Schwarzschild solution describing non-rotating black holes. Hence, predicting `$a$' of 
black holes from observed data
enables one to verify natural existence of the solution of Einstein equation 
for gravitational field of a spinning mass, more precisely the Kerr metric.

Most of the black holes are either in X-ray binaries (XRBs) or at the center of galaxies. In XRBs
they have masses at the most few tens of that of Sun, hence they are 
called stellar mass black hole (StBH). Examples of such black holes are GRS~1915+105 and GRO~J1655−40. 
The black holes at the centre of galaxies have masses atleast million times that of Sun
and are called super-massive black holes (SuBH). The black hole Sgr~$A^*$, at the centre of our
galaxy, is an example of a SuBH.  

The aim of the present paper is to predict the spin of both kinds of above mentioned black holes.
The value of $a$ is expected to vary across the black holes, which is presumably being
reflected in their various properties, e.g. quasi-periodic oscillations (QPOs), 
outflows/jets, spectra, luminosities etc. 
Currently, the major methods of measuring $a$ for black holes in XRBs are: (1) continuum spectrum fitting 
\cite{ramesh}, (2) determining the shape of the gravitationally redshifted wing of an
iron line \cite{fab}, and (3) modeling QPO of the source which is supposed to be 
linked with the black hole \cite{bm09}. For the present purpose, we adopt the third method in 
order to determine $a$ in XRBs. As the QPO frequencies seem to be scaling inversely
as mass ($M$) of the black hole, for SuBHs they are expected to be 
several orders of magnitude smaller than those in XRBs \cite{bm08} and hence it is difficult 
to identify them
in observed data. As a result, predicting $a$ is uncertain for SuBHs based on QPOs.
Therefore, we attempt to predict the spin of SuBHs, for the present purpose, based on 
underlying jet properties. In particular, we adopt a class of AGNs with jets nearly aligned to the 
observer's line of sight, viz. blazar.

In the next section, first we outline the properties of QPOs, then we model to describe 
them and finally predict spin of the black holes in XRBs. In \S 3, we briefly discuss the properties of blazars and 
then argue that the spin difference between Flat Spectrum Radio Quasars (FSRQs) and BL Lacs is 
responsible for the difference in their luminosities.
Based on that we attempt to predict possible values of 
spin of black holes in FSRQs and BL Lacs. Note that blazars are radio loud active galaxies
having very high luminosities which are expected to harbor a highly spinning
black hole generally \cite{bgm,tnm}. Finally we summarize our results in \S 4.

\section{Spin of black holes in X-ray binaries}

\subsection{Classification of QPOs}

We attempt to predict the spin of the black holes in XRBs addressing the underlying QPOs.
The QPOs in XRBs (particularly low mass XRBs) are generally of three types in terms of their 
observed frequencies ---
the ones of the order of a few hundreds to kilo Hertz: often called
high frequency QPOs (HFQPOs) for black holes and kilo Hertz (kHz) QPOs for neutron stars, 
the others of the order of 
a few tens of Hertz: intermediate frequency QPOs (IFQPOs), and finally the QPOs of the order 
of one tenth of Hertz to few tens of Hertz: low frequency QPOs (LFQPOs).

The HFQPOs/kHz QPOs 
are expected to be linked with the spin parameter of the compact object itself \cite{bm09,rezz}.
In certain black hole and neutron star systems they appear in pair, e.g. GRS~1915+105, XTE~J1550−564,
GRO~J1655−40, 4U~1636-53, 4U~1728−34, Sco~X-1. It was observed that frequencies in kHz pairs are separated 
by the order of either the spin frequency or half the spin frequency of the neutron star. Moreover,
HFQPOs in pairs appear in a $2:3$ ratio. 
However, sometimes kHz QPOs, e.g. that in Sco~X-1,
seem to appear in a $2:3$ ratio as well \cite{abra}.
All these properties favor the idea of a resonance mechanism behind the origin of HFQPOs/kHz QPOs.
The difference between HFQPOs in black holes and kHz QPOs in neutron stars might be due to 
the presence of a rotating magnetosphere in the latter case, imprinting directly the spin 
frequency on the oscillations of the disk. 
In order to model their properties, earlier 
Mukhopadhyay \cite{bm08,bm09} predicted spin of those underlying black holes and 
neutron stars, whenever unknown, for a given mass
of the compact object. 

On the other hand, LFQPOs are thought to be produced by shock oscillations in
 accretion flows \cite{msc} and, unlike HFQPOs, expected not to be related directly
with the spin of compact objects.
Of course, with the change of spin, properties of accretion flows and consequently the underlying shock
change, which can influence LFQPOs indirectly. What about IFQPOs? We will discuss them based on 
our model.

\subsection{Modeling QPOs}

Observations \cite{pal-mau} indicate a strong correlation among the QPO
frequencies in a wide range,
which supports the idea that QPOs are universal physical processes. 
It was already shown that QPOs may arise from the nonlinear hydrodynamic/hydromagnetic phenomena
of accretion flows \cite{chaos1,chaos2}, revealing them as the outcome of nonlinear resonance
(e.g. \cite{bm08,bm09} and references therein) among the various modes of the systems. 
Therefore, we argue that IFQPOs, which are
about an order of magnitude smaller than HFQPOs/kHz QPOs (sometimes, however, IFQPOs are just about 
half the corresponding HFQPOs/kHz QPOs), also result from nonlinear resonance processes in accretion flows. 

If the accretion flow with radial and vertical epicyclic modes of oscillation is excited by an external mode, 
for the present model the disturbance mode due to the rotation/spin of the compact object, then the 
frequencies of original modes and disturbance mode are 
shown \cite{bm08,bm09} to have commensurable relations which may cause the 
modes to be strongly coupled and yield an internal resonance. This leads to the 
frequencies of newly formed modes having relations 
\begin{eqnarray}
\nonumber
\nu_r+\frac{n}{2}\nu_s=\nu_z-\frac{\nu_s}{2}+\frac{m}{2}\nu_s,\\
{\rm when}\,\,\,\nu_r+\frac{n}{2}\nu_s=\nu_h;\,\,\, \nu_z-\frac{\nu_s}{2}=\nu_l,
\label{comm}
\end{eqnarray}
where $\nu_h$ and $\nu_l$ are the QPO frequencies, $\nu_r$ and $\nu_z$ are respectively 
the radial and vertical epicyclic oscillation frequencies given by
\begin{eqnarray}
\nonumber
\nu_r=\frac{\nu_o}{r}\sqrt{\Delta-4(\sqrt{r}-a)^2},\\
\nu_z=\frac{\nu_o}{r}\sqrt{r^2-4a\sqrt{r}+3a^2},
\label{epi}
\label{comm}
\end{eqnarray}
where $\nu_o$ is the orbital frequency, $\Delta=r^2-2Mr+a^2$, $r$ is the radial distance from the central
compact object, $\nu_s$ is the spin frequency of the compact object and $m$ and $n$ are integers.
For a black hole, if the excitation and then corresponding coupling take place in such a way that $m=n=2$,
then from equation (\ref{comm}), $\Delta\nu=\nu_h-\nu_l\lsim\nu_s$, with $\nu_z-\nu_r\lsim\nu_s$.
However, for a coupling with $m=n=1$, $\Delta\nu=\nu_h-\nu_l\sim\nu_s/2$, when
$\nu_z-\nu_r\lsim\nu_s/2$. 

\subsection{QPOs in GRS~1915+105 and its spin}

\begin{figure} 
\includegraphics[width=0.60\textwidth,angle=0]{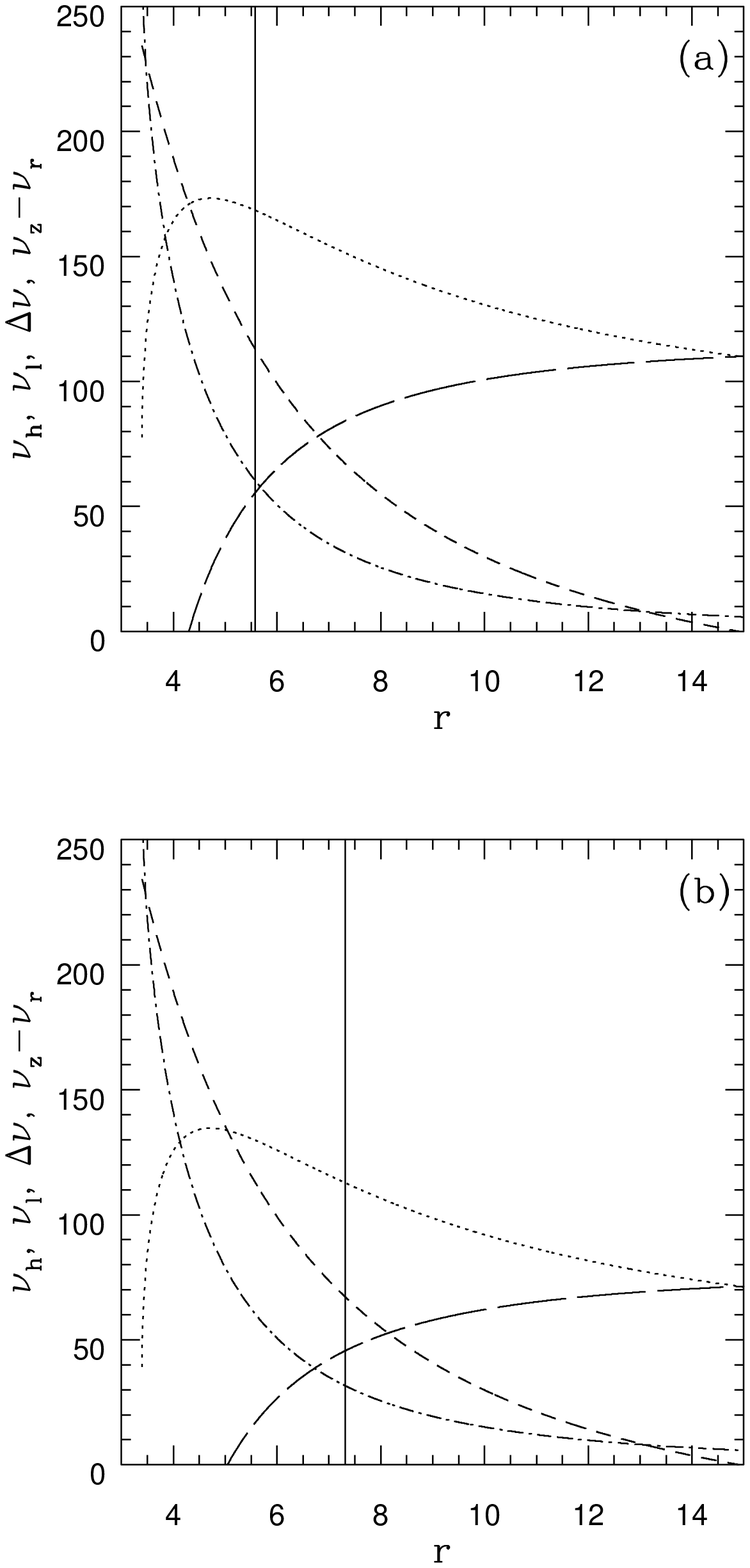}
\caption{Representative case of GRS~1915+105: (a)
Variation of frequencies of the higher (dotted line) and lower (dashed line) HFQPOs, their difference 
(long-dashed line) and difference between vertical and radial epicyclic frequencies (dot-dashed
line) in Hz as functions of radial coordinate in units of $GM/c^2$ of the accretion disk for
 $m=n=2$.
(b) Variation of frequencies of the lower (dotted line) of HFQPOs and IFQPO (dashed line), their 
difference (long-dashed line) and difference between vertical and radial epicyclic frequencies 
(dot-dashed line) in Hz as functions of radial coordinate in units of $GM/c^2$ of the accretion 
disk for $m=n=1$. The vertical solid line pinpoints the location of resonance. 
Other parameters are $M=14M_\odot$, $a=0.7$; see TABLE I for other details.
}
\label{grs}
\end{figure}

Figure \ref{grs}a shows that a pair of QPOs form for $m=n=2$
due to the coupling among the modes yielding internal resonance. We know that GRS~1915+105 exhibits 
a pair of HFQPOs and its mass is expected to be in the range of $10-18M_\odot$. Hence, we choose 
$M=14M_\odot$
in the above figure and the upper and lower HFQPOs are produced at $\nu_h=168.5$ Hz and 
$\nu_l=113$ Hz respectively for $a=0.7$, which are the observed HFQPOs for GRS~1915+105. 
Details are given in TABLE I. Note that each resonance produces two frequencies and hence IFQPO
is absent in the above process.
However, the source indeed exhibits an additional IFQPO with frequency $\nu_i=67$ Hz. Hence, a 
successful model should
reproduce this additional QPO frequency as well, for the same set of $M$ and $a$ which gives rise to the 
HFQPOs. We find that in fact the IFQPO with the observed $\nu_i$ forms in our model for the coupling with 
$m=n=1$ of the fundamental modes, as shown in Fig. \ref{grs}b revealing internal resonance, along with $\nu_l$. 
However, if we choose $M=18M_\odot$,
the extreme possible mass for GRS~1915+105, then $\nu_h,\nu_l,\nu_i$ are reproduced by our
model for $a=0.77$. See TABLE I for other details.

\begin{table}
\begin{tabular}{cccccccccc}
\hline
\hline
black hole & $M$ & $a$ & $\nu_h$ & $\nu_l$ & $\nu_i$ 
& $r_{QPO}$ & $r_{ms}$ & $n,m$ \\
\hline
\hline
GRS~1915+105 & $14$ & $0.7$ & $168.48$ & $112.95$ & $-$ & $5.58$ & $3.39$ & $2$\\
 & $14$ & $0.7$ & $-$ & $112.76$ & $67.08$ & $7.31$ & $3.39$ & $1$\\
\hline
 GRS~1915+105 & $18$ & $0.77$ & $169.59$ & $112.98$ & $-$ & $4.38$ & $3.06$ & $2$\\
 & $18$ & $0.77$ & $-$ & $115.72$ & $67.03$ & $5.81$ & $3.06$ & $1$\\
\hline
 & $18$ & $0.972$ & $-$ & $272.86$ & $69.44$ & $1.84$ & $1.71$ & $1$\\
GRS~1915+105 & $14$ & $0.975$ & $-$ & $421.63$ & $67.07$ & $2.22$ & $1.68$ & $1$\\
 & $18.4$ & $0.95$ & $167.35$ & $114.61$ & $-$ & $1.94$ & $1.94$ & $1$\\
\hline
GRS~1915+105 & $16$ & $0.62$ & $165.03$ & $111.45$ & $-$ & $4.28$ & $3.74$ & $2$\\
& $16$ & $0.62$ & $70$ & $39.28$ & $-$ & $9.8$ & $3.74$ & $1$\\
\hline
\hline
XTE~J1550-564 & $9.1$ & $0.715$ & $274.03$ & $184.19$ & $-$ & $5.3$ & $3.32$ & $2$ \\
 & $9.1$ & $0.715$ & $-$ & $175.92$ & $92.43$ & $7.51$ & $3.32$ & $1$ \\
\hline
\hline
high spinning & $15$ & $0.99$ & $659$ & unphysical & $-$ & $1.45$ & $1.45$ & $2$\\
 & $15$ & $0.99$ & $329.87$ & unphysical & $-$ & $1.45$ & $1.45$ & $1$\\
\hline
\hline
low spinning & $5$ & $0.1$ & $48.1$ & $42.8$ & $-$ & $26.81$ & $5.67$ & $2$\\
 & $5$ & $0.1$ & $36.5$ & $32.95$ & $-$ & $31.45$ & $5.67$ & $1$\\
\hline
\hline
\end{tabular}
\caption{$M$ is given in units of $M_\odot$, $\nu_{h,l,i}$ are in Hz, 
locations of QPO ($r_{QPO}$) and last stable circular orbit ($r_{ms}$) in accretion disk are
in units of $GM/c^2$. For GRS~1915+105, the observed $\nu_h=168$,
$\nu_l=113$, $\nu_i=67$ and for XTE~J1550-564, the observed $\nu_h=276$, $\nu_l=184$,
$\nu_i=92$ (controversial).
}
\label{tab2}
\end{table}

Narayan and his collaborators, based on fitting the continuum spectra, argued \cite{ramesh} that
$a>0.98$ for GRS~1915+105. Therefore, in our work, first we tried to confirm a large spin 
for this black hole. It, however, appears that $a=0.975$ and $0.972$, clearly $<0.98$, to reproduce an IFQPO
for $M=14M_\odot$ and $18M_\odot$ respectively for $m=n=1$. 
Then for the pair of HFQPOs, our model predicts $a=0.95$, again $<0.98$, for the same $m$ and $n$ at 
a choice of rather 
higher mass $M=18.4M_\odot$ \cite{bm08,bm09}. Hence, not only $a<0.98$, there is a clear mismatch between
the sets of input parameters producing observed HFQPOs and IFQPO, questioning a high spin of
GRS~1915+105. Indeed, it was shown \cite{wog} recently that the prediction of spin 
based on diskoseismology and $g$-mode oscillation does not always tally with that
from continuum fitting. Although for GRS~1915+105 the predictions from these two methods rather tally,
for other sources there are problems.

Sometimes, another QPO frequency $\sim 41$ Hz is reported 
for this source. If the set of frequencies $41$ Hz and $67$ Hz is considered to form
a pair, similar to that having $113$ Hz and $168$ Hz, then both the pairs should be 
reproduced by the same model. Assuming the first pair being the result of coupling with $m=n=1$
and the second being with $m=n=2$, they are reproduced for $M=16M_\odot$
and $a=0.62$ in our model, which indicates the spin of GRS~1915+105 to be even slower. See TABLE I for other parameters.

\begin{figure} 
\includegraphics[width=0.60\textwidth,angle=0]{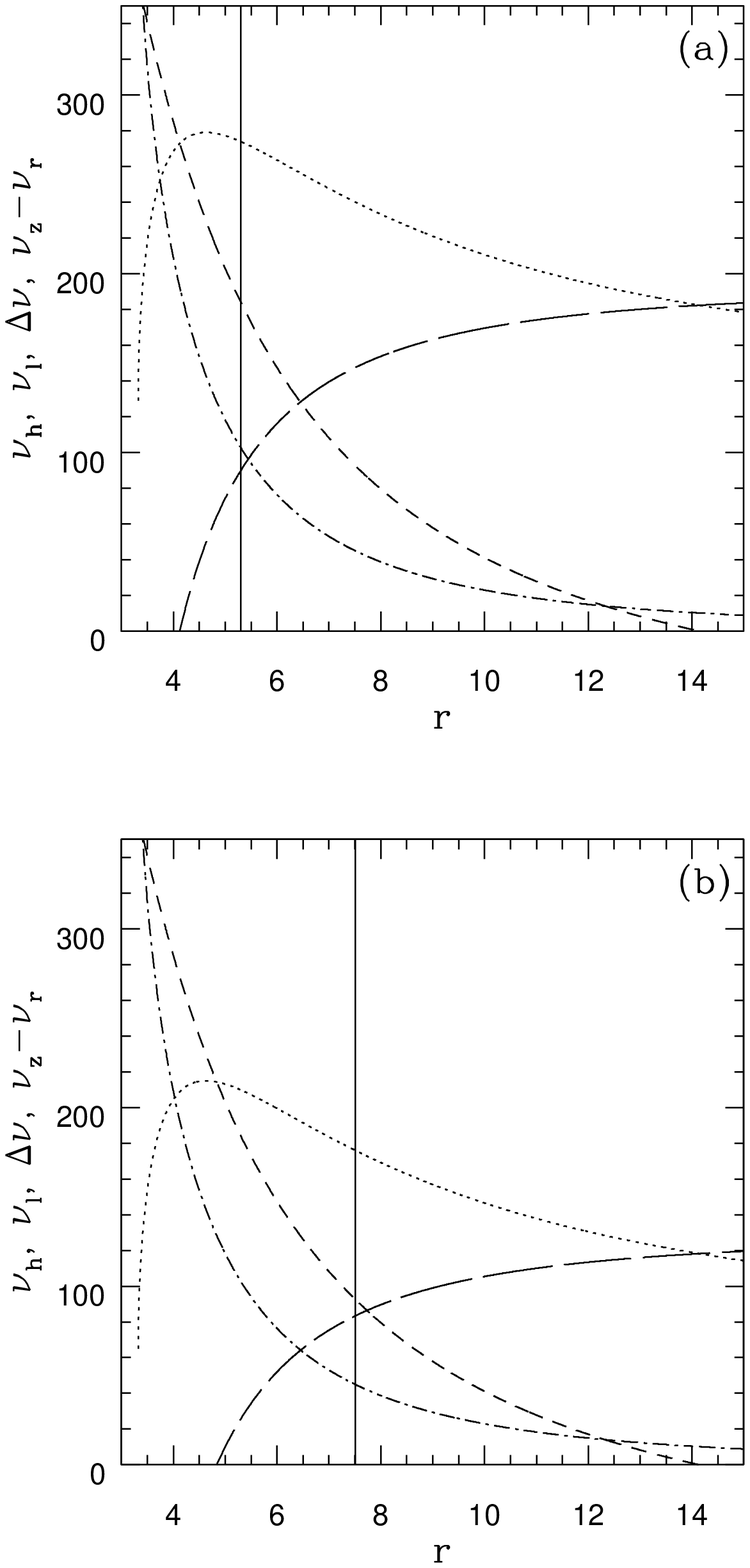}
\caption{
Representative case of XTE~J1550-564: (a)
Variation of frequencies of the higher (dotted line) and lower (dashed line) HFQPOs, their difference 
(long-dashed line) and difference between vertical and radial epicyclic frequencies (dot-dashed
line) in Hz as functions of radial coordinate in units of $GM/c^2$ of the accretion disk for
 $m=n=2$.
(b) Variation of frequencies of the lower (dotted line) of HFQPOs and IFQPO (dashed line), their 
difference (long-dashed line) and difference between vertical and radial epicyclic frequencies 
(dot-dashed line) in Hz as functions of radial coordinate in units of $GM/c^2$ of the accretion 
disk for $m=n=1$. The vertical solid line pinpoints the location of resonance. 
Other parameters are $M=9.1M_\odot$, $a=0.715$; see TABLE I for other details.
}
\label{xte}
\end{figure}
\subsection{QPOs in XTE~J1550-564 and its spin}

The black hole source XTE~J1550-564 also shows a pair of HFQPOs at $276$ Hz and $184$ Hz along with an IFQPO
at $\nu_i=92$ Hz \cite{rem}. However, the IFQPO for this source is controversial.
As the mass of the source ($M=9.1M_\odot$) is also quite definite, we can test
our model for this source in order to determine the spin of the black hole. In order
to reproduce HFQPOs and IFQPO by the couplings with $m=n=2$ and $m=n=1$ respectively, as shown
by Fig. \ref{xte},
the spin of the source is found to be $a=0.715$. Details are given in TABLE I.

\begin{figure} 
\includegraphics[width=0.60\textwidth,angle=0]{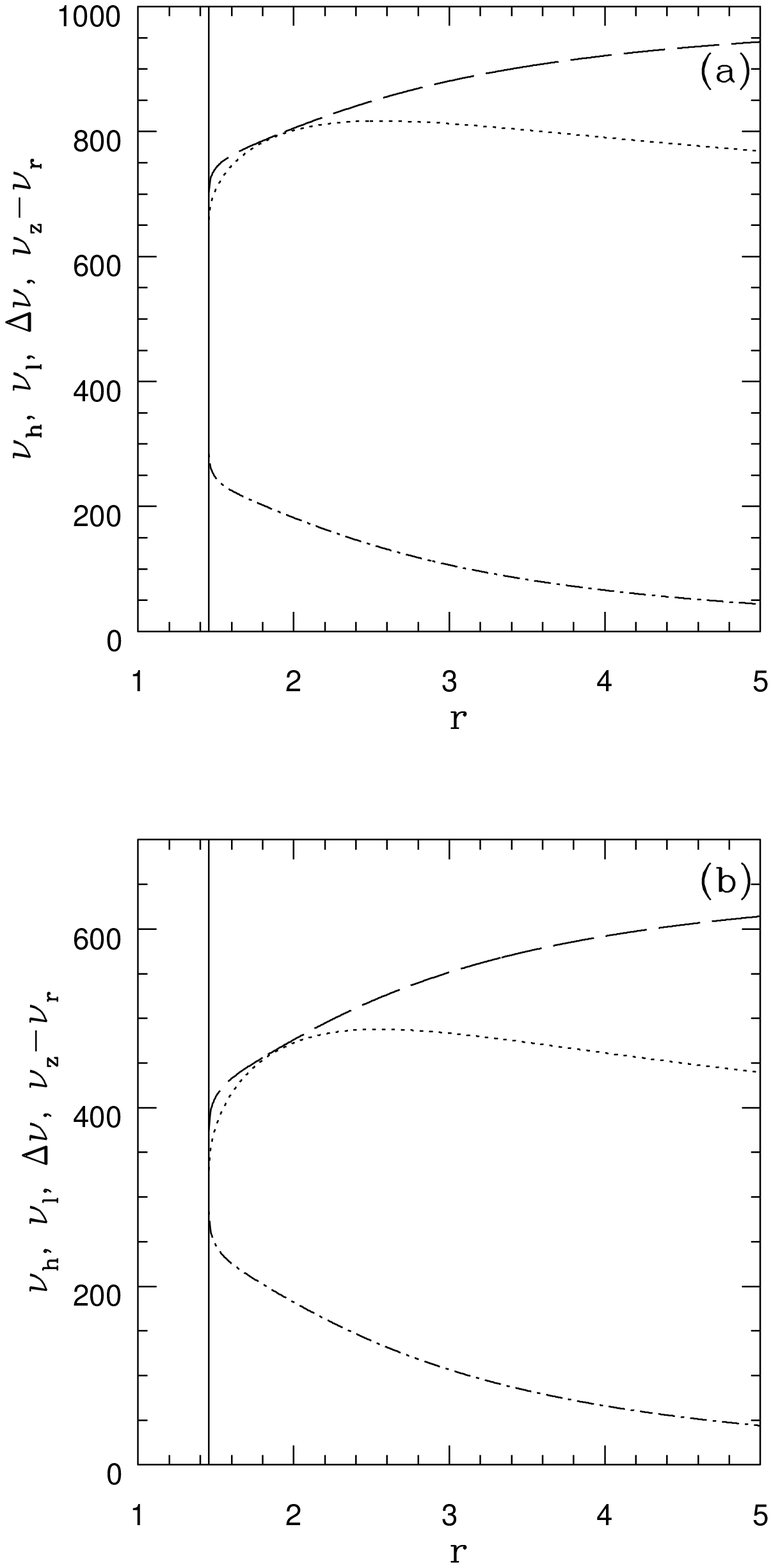}
\caption{
Representative case of a high spinning black hole with $a=0.99$: (a)
Variation of frequencies of the higher (dotted line) of HFQPOs, the difference between
higher and lower (unphysical) frequencies of HFQPOs (long-dashed line) and difference between 
vertical and radial epicyclic frequencies (dot-dashed
line) in Hz as functions of radial coordinate in units of $GM/c^2$ of the accretion disk for
 $m=n=2$.
(b) Same as in (a), but for $m=n=1$. The vertical solid line pinpoints the location of resonance.
$M=15M_\odot$; see TABLE I for other details.
}
\label{highs}
\end{figure}

\subsection{QPOs in fast spinning black holes}

Our model also predicts that a black hole with a larger spin is prone to exhibit
single HFQPO based on a nonlinear resonance coupling with $m=n=1$. Figure \ref{highs} shows a hypothetical
case with $a=0.99$ exhibiting, for $m=n=1$, $\nu_h\sim 330$ Hz and a negative and
hence unphysical $\nu_l$. Similarly, single HFQPO forms for $m=n=2$, except with a 
larger $\nu_h$, which is very similar to that of a neutron star. Hence, a rapidly spinning 
black hole (and neutron star)
seems to not favour the formation of a pair of HFQPOs. This is more unfavorable
due to the fact that a fast spinning compact object presumably creates stronger disturbance
in the surrounding disk, rendering only one QPO frequency dominant.

\begin{figure} 
\includegraphics[width=0.60\textwidth,angle=0]{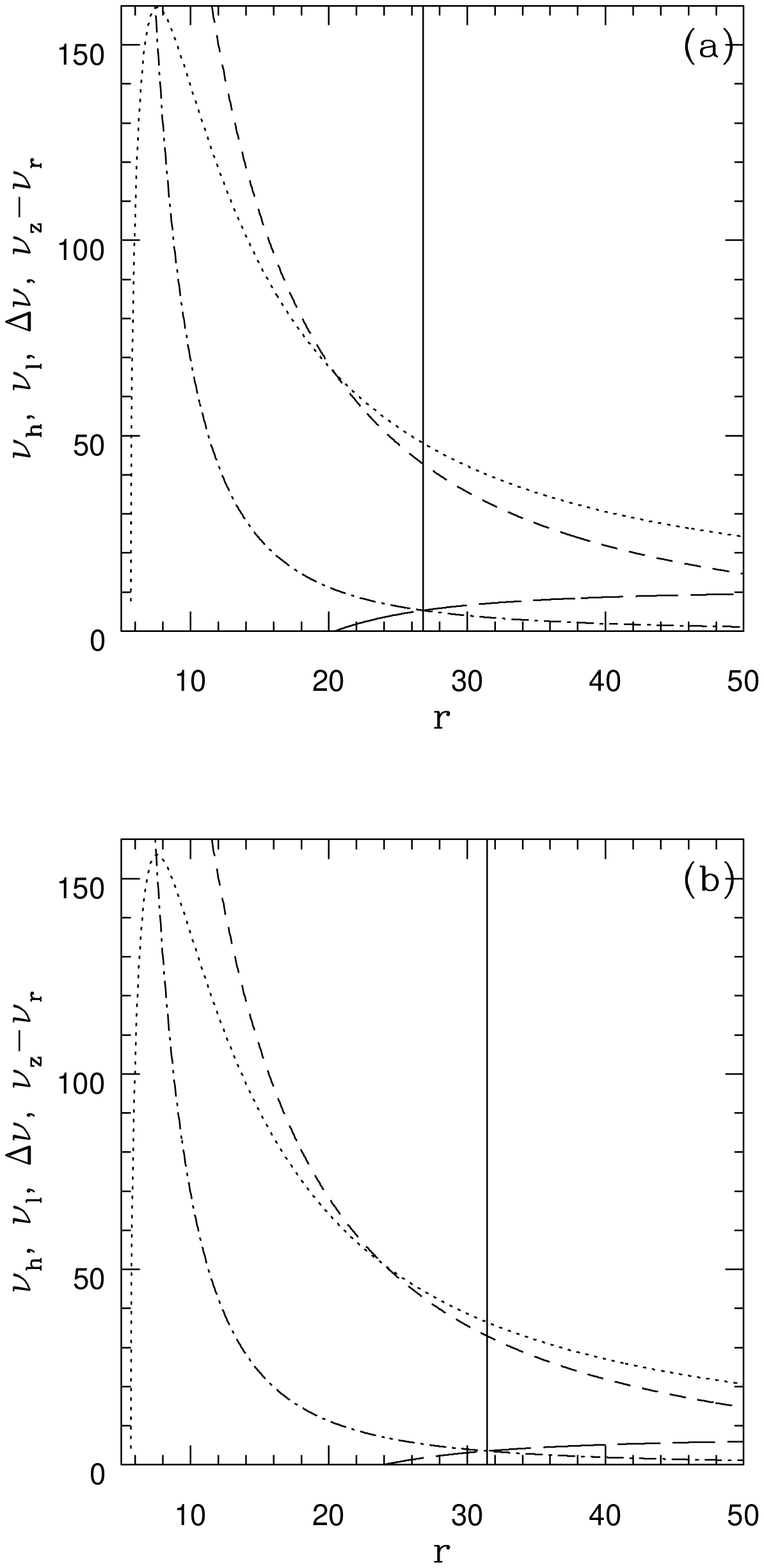}
\caption{
Representative case of a low spinning black hole with $a=0.1$: (a)
Variation of frequencies of the higher (dotted line) and lower (dashed line) HFQPOs, their difference 
(long-dashed line) and difference between vertical and radial epicyclic frequencies (dot-dashed
line) in Hz as functions of radial coordinate in units of $GM/c^2$ of the accretion disk for
 $m=n=2$.
(b) Same as in (a), but for $m=n=1$.
The vertical solid line pinpoints the location of resonance. 
$M=5M_\odot$; see TABLE I for other details.
}
\label{lows}
\end{figure}

\subsection{QPOs in slow spinning black holes}

It is easy to check that for a low spinning black hole, the results with $m=n=1$ and $m=n=2$
(almost) coincide giving rise to low values of QPO frequency when $\nu_h\sim\nu_l$. 
Figure \ref{lows} shows that for $a=0.1$, resonance occurs at a larger radius
away from the black hole where $\nu_r$ merges to $\nu_z$ with $\nu_s\rightarrow 0$.
This is justified as, a smaller spin corresponds to a smaller strength of disturbance,
which results in practically no nonlinear effects in the system, when the linear effect
is also very weak. Hence, practically the coupling and hence the resonance takes place 
among the fundamental modes of the system --- due to the radial and vertical epicyclic oscillations. 
This further implies that HFQPOs can not arise in low spinning black holes.

\section{Spin of black holes in Blazars}

\subsection{Classification of blazars}

\begin{figure} 
\includegraphics[width=0.90\textwidth,angle=0]{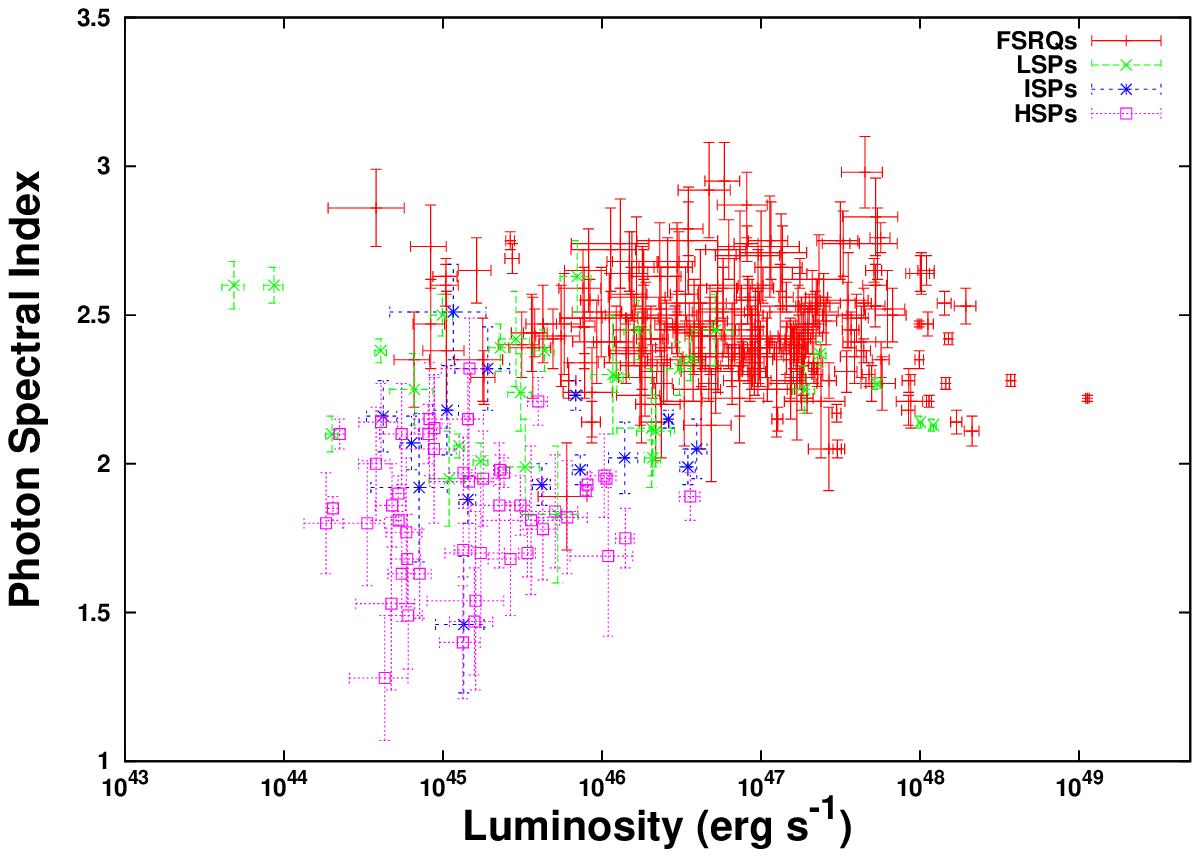}
\caption{
Distribution of observed luminosity and spectral index for FSRQs and BL Lacs. Red points 
represent FSRQs, green, blue and violet points respectively represent LSPs, ISPs and HSPs. 
}
\label{obs}
\end{figure}

Here we look at the issue of spin addressing the underlying luminosities.
Blazars are the radio loud AGNs, which are divided into two classes: FSRQ and BL Lac.
FSRQs are considered to be the subclass of FR II galaxies, whereas BL Lacs to be the subclass of 
FR I galaxies, based on the unification scenario \cite{urry1995,antonucci1993}. Radio 
observations suggest that the jets of FR II galaxies are more collimated and more 
powerful than that of FR I galaxies. 
During its first year of observation, the Large Area Telescope (LAT) on
board Fermi Gamma-ray Space Telescope (Fermi) detected around $600$ blazars ---
almost an equal number of 
FSRQs ($281$) and BL Lacs ($291$) \cite{abdo1,abdo2}. 
Based on their spectral energy distribution (SED), BL Lacs are
further classified into three sub-categories: the low synchrotron peak
(LSP), the intermediate synchrotron peak (ISP) and the high synchrotron
peak (HSP) sources \cite{abdo3}. With this significantly improved
catalog, one can for the first time carry out a detailed comparison of 
${\gamma}$-ray properties of FSRQs and BL Lacs and try to understand the 
underlying physics related to their intrinsic luminosities.
Note that observationally FSRQs are on average about three orders of magnitude more luminous than BL Lacs. 
Also, BL Lacs exhibit a harder average photon spectrum 
than that of FSRQs and the luminosities of LSP sources are closer to the 
luminosities of FSRQs. From ${\gamma}$-ray spectral studies, it is observed 
that usually FSRQs show a break in the spectrum, while among BL Lacs, mostly 
LSPs show such a feature \cite{abdo3}. Therefore, we consider FSRQs 
and LSPs as one group (FSRQ group) and ISPs and HSPs as another 
group (BL Lac group). In rest of the paper, we will use FSRQ group and FSRQs
interchangeably and same for BL Lac group and BL Lacs. Figure \ref{obs} confirms that FSRQs exhibit
on average three orders of magnitude higher observed luminosity and larger photon spectral
index than those of BL Lacs. Moreover, FSRQs show strong emission lines, 
whereas BL Lacs do not.

\subsection{Properties of blazars}

Based on the isotropic luminosity and spectral index of these sources, earlier
authors \cite{ghisellini2009}  proposed that FSRQs accrete at or
above a critical mass accretion rate ($\dot{M}_{cric}$), whereas
BL Lacs at a rate below $\dot{M}_{cric}$, which results in their
observed large luminosity differences. However, as mentioned as well by others \cite{xu},
the actual cause of the FR I/FR II (broader classes of FSRQs/BL Lacs, as mentioned above) 
division is still not confirmed. There are two models.
One argues that the interaction of jets with the ambient medium of different physical properties 
causes morphological differences between them (e.g. \cite{gopal}). The other argues based on
different intrinsic nuclear properties of accretion and
jet formation processes (e.g., \cite{baum}) between them. 

Indeed, spectral energy distribution (SED) modeling suggests that jet emissions of FSRQs are dominated by 
external Compton (EC) processes.
In this case, the relativistic particles in the jet 
upscatter external low-energy photons leading to the generation of ${\gamma}$-ray photons. The
source of low-energy photons could be the underlying accretion disk, the broad-line region,
the dusty torus etc. \cite{dermer,dermer2,marcher,poutanen,ghise,sikora,mal,baej,fos,tave},
depending on how close the gamma-ray emitting region
is to the black hole. On the other hand, SEDs of BL Lac sources are better fitted by synchrotron 
self-Compton (SSC) emissions, when
the synchrotron photons of the jet are
upscattered via inverse Compton processes by relativistic particles in the jet. Therefore, 
observed data suggest that the 
relativistic beaming effects in the two classes are not same and hence the 
intrinsic luminosity (which is the luminosity corrected for
relativistic beaming), which is the property of the underlying accretion disk, should be treated as 
a more fundamental quantity than the observed luminosity in order
to understand the origin of difference between the two source classes. However, if the source classes
exhibit different accretion rates, intrinsic luminosities (i.e. that in the disk and hence at the base of the jet) 
also have to be different from each other.

Mukhopadhyay and his collaborators \cite{bgm} 
showed that the black hole spin plays a crucial role to control outflow and corresponding power. 
They considered a disk-outflow coupled region and by solving the hydrodynamic 
equations found that outflow is stronger for a faster rotating black hole 
than that for a slower one. 
Therefore, even with the 
same accretion rate, one can have a difference in luminosities due to the 
difference in the spin of central black hole. 

\subsection{Difference between intrinsic and observed luminosities}

It is assumed that the blazar luminosity mostly comes from the jet which is
believed to be made up of relativistic particles, either 
in a continuous flow or as discrete blobs. The $\gamma$-ray emission 
of jet arises due to
either SSC and/or EC processes, as mentioned above.
The observed beamed
luminosity ($L_{obs}$) and the jet frame luminosity ($L_{unbeamed}$) are related by \cite{dermer1995}
\begin{equation}
L_{obs} = L_{unbeamed}~\delta^{m+n},
\label{beam}
\end{equation}
where $m=2$ for a continuous jet and $m=3$ for a discrete jet, $n={\alpha}_{\gamma}$ for SSC process and 
$n=2{\alpha}_{\gamma}+1$ for EC process, when $\alpha_\gamma$ is the energy spectral index. 
One can obtain the intrinsic luminosity $L_{unbeamed}$ of the source from equation (\ref{beam}) 
by knowing $L_{obs}$ and $\delta$.
We use an average value of $\delta$ for blazars and individual spectral 
indices while calculating unbeamed/intrinsic luminosities of FSRQs and BL Lacs.
In order to obtain $L_{unbeamed}$, we consider a continuous jet for
the BL Lac group having SSC emission. However, for the FSRQ group, we consider a combination of SSC and EC 
emission models in the framework of continuous jet. 
Since the fraction of SSC and EC combination is not well constrained, we consider different 
scenarios: (a) $50\%$ SSC and $50\%$ EC contributions,
(b) $25\%$ SSC and $75\%$ EC contributions, (c) $10\%$ SSC and $90\%$ EC contributions, and (d)
 $100\%$ EC contribution. 


\subsection{Fixing $\delta$ in order to determine intrinsic luminosities}

A very important factor to determine the intrinsic luminosity of a blazar is $\delta$.
There are mainly two approaches to determine $\delta$, actually Lorentz factor $\Gamma$ which 
is related to $\delta$: 
(i) SED modeling, (ii) radio Very Large Baseline Interferometry (VLBI) measurement.
Considering the opacity arguments, it is generally believed that the $\gamma$-ray emitting region is located 
closer to the central black hole. 
Hence, SED modeling beyond radio may be more appropriate, since it more accurately represents the region of 
$\gamma$-ray emission.
Unfortunately, in most cases SED data gathered across wavelengths are not simultaneous. 
Further, SED modeling involves a large number of model parameters and, hence, typically results in large 
uncertainties. 
Another important parameter, jet to line-of-sight angle ($\theta_{jet}$), is not measured, 
rather a nominal value of $\theta_{jet}$ is typically considered for this modeling. 
Moreover, for most of the cases, the goodness of fit is not examined and, hence, the parameter values are not 
optimized. 
Based on SED modeling of $85$ blazars from the first $3$ 
months' Fermi observation, it was found \cite{ghisellini2010} that the values of $\Gamma$ for blazars are 
within the range of $10-15$. 

The values of $\Gamma$ and $\delta$ can also be determined from VLBI observations based on 
direct observation of the brightness temperature of the 
source ($T_{b,obs}$). Comparing $T_{b,obs}$ with the intrinsic brightness temperature of 
the source ($T_{b,int}$), one can determine $\delta$ \cite{hovatta2009,savolainen2010}. 
From the observed apparent velocity ($v_{ap}$) and $\delta$, $\theta_{jet}$ and $\Gamma$ can be derived. 
However, in this technique $T_{b,int}$ is often assumed to be the equipartition temperature. 
It was considered in earlier works \cite{hovatta2009, savolainen2010} that $T_{b,int}$ is 
same for all sources. This assumption makes the estimated values of $\Gamma$ and $\delta$ less accurate. 

The values of $\Gamma$ and $\delta$ could also be determined \cite{jorstad2005}
from VLBI observations adopting yet another technique. 
In this technique, $\delta$ is derived by comparing the timescale of decline in flux density with the 
light travel time across the emitting region. From the knowledge of $\delta$ and $v_{ap}$, 
the previous authors \cite{jorstad2005} derived 
$\theta_{jet}$ and $\Gamma$. 
This approach yields more appropriate values in comparison to other alternate methods. In order 
to determine the average $\delta$ and $\Gamma$ from this approach \cite{jorstad2005}, we assume that the intrinsic
$\delta$ (also $\Gamma$) distribution and the corresponding errors are Gaussian (following 
the methodology used in \cite{venters2007} and \cite{bsm} 
to determine the average spectral index of blazars). The average $\delta$ for FSRQs is then obtained to be 
$23.1 \pm 8.9$ and that for BL Lacs to be $15.3 \pm 5.5$ 
and the respective values of $\Gamma$ to be $17.2 \pm 5.3$ and $12.5 \pm 3.5$. 
For the present work, we choose $\delta$ and $\Gamma$ obtained from this approach.
Considering the large errors, we adopt the average values of $\delta = 20.6 \pm 8.4$ and $\Gamma = 15.1 \pm 4.6$ for both the source classes. 

There is, however, a concern regarding the difference in location of $\gamma$-ray emission versus the 
location of radio emission.
However, the independent estimates of $\Gamma$ values derived from the SED modeling \cite{ghisellini2010} lie in 
the range $10-15$, which is consistent with the mean value 
adopted here. This probably indicates that the parameters $\delta$ and $\Gamma$ are not 
varying significantly between the two regions. We use the average value of $\delta=20.6$ to estimate the 
unbeamed luminosities.

\begin{figure} 
\includegraphics[width=0.90\textwidth,angle=0]{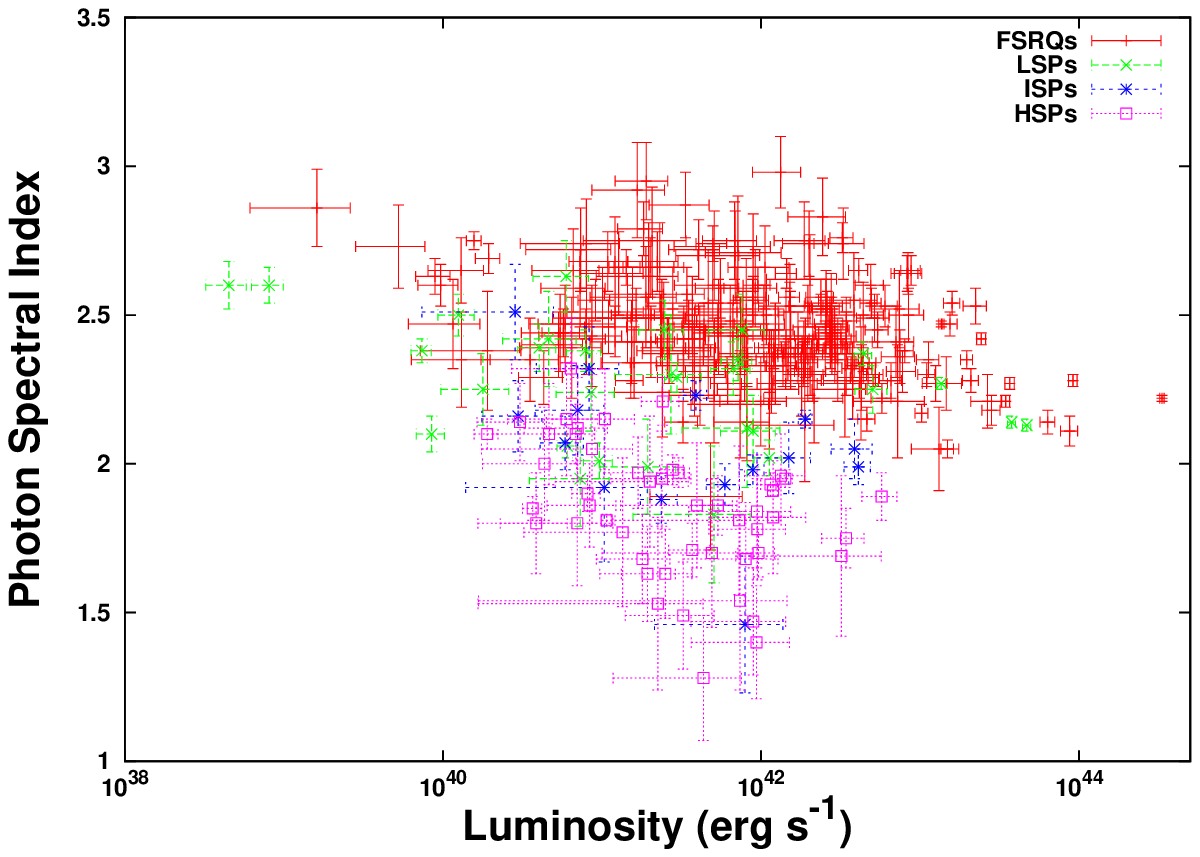}
\caption{
Same as in Fig. \ref{obs}, except intrinsic/unbeamed luminosities are plotted in the horizontal axis.
Contributions from $50\%$ SSC and $50\%$ EC are chosen for FSRQs.
}
\label{intrin}
\end{figure}

\subsection{Evaluating intrinsic luminosity and its dependence on the spin of black hole}

With the above $\delta$, using equation (\ref{beam}), interestingly the unbeamed/intrinsic luminosity difference between FSRQs 
and BL Lacs is shown in Fig. \ref{intrin} to decrease significantly 
in comparison to the observed values.
There is, however, a small difference in intrinsic luminosities between
BL Lacs and FSRQs (with FSRQs on average having higher luminosity). 
Nevertheless, the small difference in intrinsic luminosities 
suggests that it is difficult to accommodate models 
involving vastly different accretion rates in 
these two systems. Any change in accretion rate would affect the properties of the 
flows in the accretion disk itself, before it is launched and beamed.
We propose that the intrinsic luminosity mismatch arises from the difference in the spin of the central black holes. 
Indeed one expects some connection between the jet 
and the central black hole \cite{blandford1977,nm}.
In an earlier paper \cite{bgm}, 
based on a disk-outflow coupled model,
the fundamental properties of central black hole were shown to be linked with accretion and outflow,
which in turn could be related to the properties of FSRQs/BL Lacs.
In that work, the total mechanical outflow luminosity was found to increase
with the increase of spin of the central black hole. This implies,
for a faster spinning black hole, matter will flow out faster,
making the outflow stronger with higher flow density.

For a fixed accretion rate, it is known (see e.g. \cite{rajesh2010,mg2003},
for the latest solutions) that with the increasing spin
of black hole, the optically thick part of the accretion disk
advances towards the black hole. Therefore, the supply of soft photons to the jet emitting
region increases with the increasing spin of black hole.
On the other hand, as shown earlier \cite{bgm}, with the increase of spin,
density of the disk-outflow coupled region at a particular
accretion rate increases.
Therefore, with the increase of spin, denser matter having larger
supply of soft photons will interact with radiation more rapidly,
which results in losing more energy. Hence matter in the jet
and then corresponding radiation is expected to exhibit a steeper
spectrum. 
This agrees with the observed steeper ${\gamma}$-ray
spectrum of FSRQs than that of BL Lac objects. 
It was found earlier \cite{ledlow}, based on the host galaxy optical
luminosity---radio luminosity plane dividing FR I and FR II radio galaxies,
that radio power is proportional to the optical
luminosity of the host galaxy. However, as discussed in \S III.D that 
basic properties in radio and $\gamma$-ray emitting regions do not vary significantly. 
Hence, all the above properties also explain 	
the existence of strong ionizing continuum in FSRQs, which are expected to harbor faster spinning 
black holes (as explained further in \S III.F below) and therefore exhibit denser accretion flow, revealing
strong emission lines in them.

\subsection{Spin difference between black holes in FSRQs and BL Lacs}

It is also very important to note that although 
the unbeamed luminosities of FSRQs and BL Lacs are much smaller than their observed values
(and the difference of unbeamed luminosities between them decreases significantly, 
as described above), FSRQs 
still exhibit higher luminosities than BL Lacs. This 
implies that FSRQs harbor fast spinning black holes than BL Lacs \cite{bgm}.
Note that it is unreasonable to believe that there is a $100\%$ EC contribution
in observed luminosities of FSRQ group, which will make their intrinsic
luminosities and hence spin to be smaller than that of BL Lacs --- that goes against their
spectral behavior as mentioned above.
Therefore, we consider the bound on the contribution from EC process to be $\lsim 95\%$.

If one considers that the magnetic 
field plays a major role in collimating the jets \cite{blandford1977,mckinney2009}, 
it can result in more twisting of magnetic field lines and, hence, a more collimated jet 
for a faster rotating black hole, lending further support for enhanced spin in FSRQ systems.
By numerical simulations, it has been already investigated \cite{tnm} that whether the spin of central
black hole can play a major role to explain radio loud/quiet dichotomy or not. 
For a geometrically thick disk, as is also considered in \cite{bgm}, simulations show that the 
choice of a spin value 
$a{\sim}0.15$ of a black hole for radio quiet AGNs can explain their difference in 
luminosities from radio loud AGNs having a maximally spinning black hole. 

\subsection{Predicting spin for black holes in FSRQs and BL Lacs}

\begin{figure} 
\includegraphics[width=0.60\textwidth,angle=0]{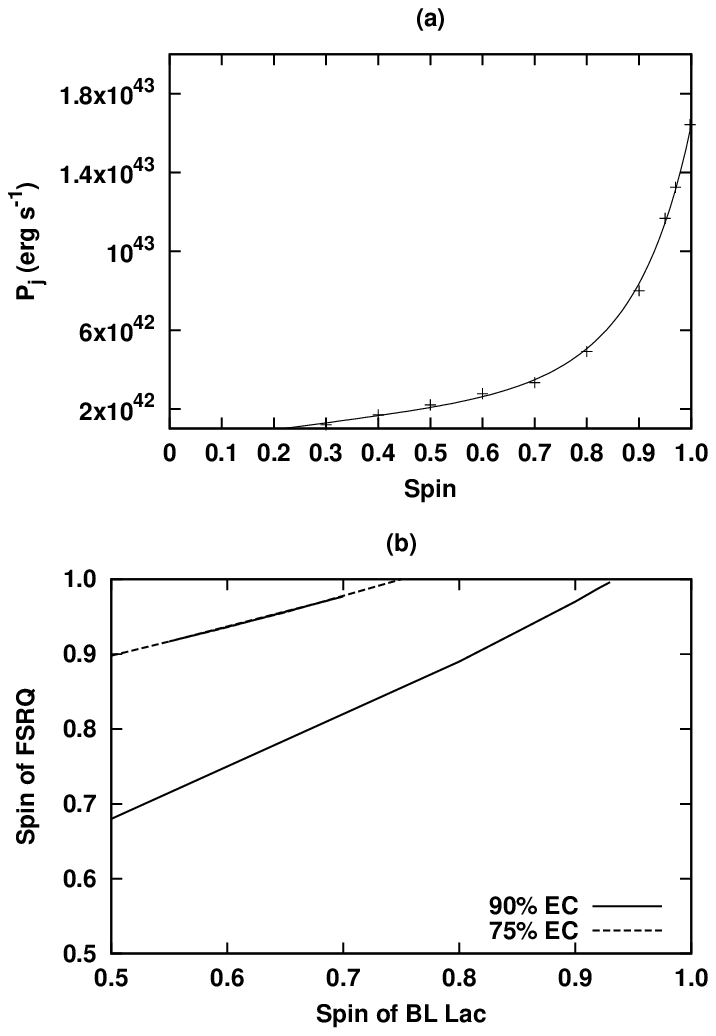}
\caption{
(a) Total mechanical power of the outflow $P_j$ (points) as a function of spin (from \cite{bgm}), when
solid line represents the fit with the function given by equation (\ref{pa}).
(b) The values of FSRQ spin for a range of BL Lac spin. We take out the beaming effect considering 
SSC emission in BL Lacs and a combination of SSC and EC emissions in FSRQs.}
\label{powers}
\end{figure}

Now recalling the relationship connecting the total mechanical power of outflow ($P_j$) with the spin of black 
hole given by equation (20) and shown in Fig. 7(a) of \cite{bgm}, we derive a simple relationship between $P_j$ and $a$,
as shown by Figure \ref{powers}(a),
based on fitting the power given by the above mentioned equation in \cite{bgm} by a third order polynomial of $a$ as
\begin{equation}
P_j =10^{Aa^3+Ba^2+Ca+D},
\label{pa}
\end{equation}
where $A= 2.87{\pm}0.26$, $B= -4.08{\pm}0.40$, $C= 2.88{\pm}0.17$ and $D= 41.53{\pm}0.02$. 
Since the earlier work \cite{bgm} did not consider any beaming effect ($\Gamma = 1$) while calculating 
$P_j$, which is calculated in the frame of disk-outflow coupled system, we assume in the first approximation 
that $P_j$ is proportional to the intrinsic luminosity such that 
\begin{eqnarray}
\frac{P_{j,FSRQ}}{P_{j,BLLac}} = \frac{L_{unbeamed,FSRQ}}{L_{unbeamed,BLLac}}, 
\end{eqnarray}
but not the observed luminosity,
due to the inadequate knowledge of the underlying radiation hydrodynamics therein. 

Now, for a particular spin of the black hole in BL Lac, we calculate $P_{j,BLLac}$ using equation (\ref{pa}). 
Since the ratio of $P_j$ of the two systems is equal to 
the ratio of their intrinsic luminosities, the corresponding power of FSRQ ($P_{j,FSRQ}$) can be 
estimated by
knowing the average intrinsic luminosities of the two classes. 
Therefore, the corresponding average spin of the black hole in FSRQ is derived from equation (\ref{pa}). 
We compute the average
spin of FSRQs' black hole for a range of the average spin of BL Lacs' black hole, as shown in Figure \ref{powers}(b).
FSRQs and BL Lacs, being radio loud AGNs, are expected to be harboring fast rotating black holes than the
radio quiet AGNs \cite{tnm}.
Hence, assuming that the spin of black holes in radio loud AGNs (and hence of BL Lacs, which are 
argued above to be harboring slower black holes among blazars) should not be less
than $0.5$, the minimum possible spin for the black hole in FSRQs is estimated to be ${\sim}0.9$, for a case 
where the SEDs of 
FSRQs are assumed to have $75\%$ EC contribution (and ${\sim}0.68$ assuming $90\%$ EC contribution).
Conversely, assuming FSRQs to be harboring maximally spinning black holes, the maximum possible spin for the
black hole in BL Lacs is 
${\sim}0.75$, considering $75\%$
EC contribution (and ${\sim}0.93$ for $90\%$ EC contribution). 

\section{Discussion and conclusion}

By simultaneously modeling three classes of QPOs of the black hole in XRBs --- a pair of HFQPOs and an IFQPO,
we predict the spin of black holes. While the model can predict spin by fitting either
a pair of HFQPOs or an IFQPO (for black holes exhibiting only one of the classes),
the result will be robust for the black holes showing both classes of QPO. The sources
GRS~1915+105 and XTE~J1550-564 are the ones exhibing both HFQPOs and IFQPO and we predict
their spins. Our model predicts that GRS~1915+105 could be fast spinning ($a=0.95-0.972$), only if
its mass $M\gsim 18M_\odot$. But this set of mass and spin could not predict the
existence of HFQPOs along with IFQPO. For the existence of both HFQPOs and IFQPO, our model
argues for $a=0.77$, for the choice of $M=18M_\odot$. 
For $M=14M_\odot$, $a$ is obtained to be $0.7$, which is a more robust result. 

The present model also argues that a fast spinning black hole cannot exhibit a pair of HFQPOs;
it reveals only one HFQPO. Moreover, a slow spinning black hole cannot exhibit
any HFQPO (and presumably IFQPO). Hence, the pair of HFQPOs in a $2:3$ ratio seems to be the property 
of a black hole spinning neither very fast nor slow. Same could be true for neutron stars exhibiting
kHz QPOs.

In principle, above ideas could be applied for SuBHs in oder to determine their spin.
Unfortunately, due to their very low expected frequencies, it is very difficult to
probe their QPOs. Hence, in order to predict the spin of SuBHs, we model the sources,
for the present purpose blazars, differently.

We propose that there is a difference in the spin of the central black holes in the
two classes of radio loud AGNs, namely, FSRQs and BL Lacs. 
The spin difference between these two classes of radio loud AGNs is 
predicted based on their difference in the mechanical outflow powers 
(and hence, the unbeamed/intrinsic luminosities). 
We predict that FSRQs are
expected to harbor a faster spinning black hole than BL Lacs.
By the same argument, the black hole spin in LSPs is expected to be higher than that in ISPs, 
while that in HSPs to be the slowest. However, both the source classes must be harboring  
spinning black holes.

Unfortunately, due to lack of proper data required for this modeling, we
could not predict the spin of individual blazars. We rather attempt to predict the average
spin of FSRQs and BL Lacs from the large sets of respective data and 
show that general relativistic effects of rotating black hole play important role
in these systems. 

Note that recently a new type of jets was observed in the
sources of SuBHs due to tidal disruption of stars \cite{zhang1},
which are supported only by the existence of rapidly
spinning black holes at the center \cite{zhang2}.

More detailed observations of these systems in future will provide a 
potential laboratory to investigate the underlying general relativistic processes of these sources.

Therefore, by modeling observed data from XRBs and blazars based on the principle of general relativity,
we predict the spin of the black holes in the respective sources to be nonzero. This
provides a natural proof for the existence of Kerr metric.

{\bf Acknowledgments:} This work was partly supported by an ISRO grant ISRO/RES/2/367/10-11. 
Thanks are due to Upasana Das for carefully reading the manuscript.

\end{document}